\documentclass[journal]{IEEEtran}

\usepackage{algorithm}
\usepackage{algorithmicx}
\usepackage[noend]{algpseudocode} 

\usepackage[colorlinks=true, allcolors=blue]{hyperref}

\usepackage{graphicx}
\usepackage{cite}
\usepackage{times}
\usepackage{epsfig}
\usepackage{amsmath}
\usepackage{amssymb}
\usepackage{mwe}
\usepackage{acro}
\usepackage{amssymb}
\usepackage{xcolor,colortbl}
\usepackage{tabularx}
\usepackage{relsize}
\usepackage{pifont}
\usepackage{booktabs} 
\usepackage{multirow}
\usepackage{multicol}
\usepackage{adjustbox}
\usepackage{float}

\usepackage{cite}
\usepackage{acro}
\usepackage{graphicx}
\usepackage{float}
\usepackage{xcolor}
\usepackage{amsmath}

\usepackage{multirow}
\usepackage{booktabs}

\usepackage{tikz}
\def\checkmark{\tikz\fill[scale=0.4](0,.35) -- (.25,0) -- (1,.7) -- (.25,.15) -- cycle;} 

\DeclareAcronym{CNN}{
short=CNN,
long=convolutional neural network,
}

\begin{document}
%
\title{Omni-Seg: A Scale-aware Dynamic Network for Renal Pathological Image Segmentation}
%
%
%

\author{Ruining~Deng,
        Quan~Liu,
        Can~Cui,
        Tianyuan~Yao,
        Jun~Long,
        Zuhayr~Asad,
        R.~Michael~Womick,
        Zheyu~Zhu,
        Agnes~B.~Fogo,
        Shilin~Zhao,
        Haichun~Yang,
        Yuankai~Huo,~\IEEEmembership{Senior Member,~IEEE}
\thanks{*Y. Huo is the corresponding author, e-mail: yuankai.huo@vanderbilt.edu}
\thanks{R. Deng, Q. Liu, C. Cui, T. Yao, Z. Asad, Z. Zhu, Y. Huo were with the Department of Computer Science, Vanderbilt University, Nashville, TN, 37215, USA, }
\thanks{J. Long was with Big Data Institute,  Central South University,  Changsha, 410012, China}
\thanks{A. Fogo, S. Zhao and H. Yang were with the Department
of Pathology, Microbiology and Immunology, Vanderbilt University Medical Center, Nashville,
TN, 37215, USA}
\thanks{R. Womick was with the Department
of Computer Science, The University of North Carolina at Chapel Hill, Chapel Hill,
NC, 27514, USA}
}

\markboth{Manuscript pre-print, Jan~2023}%
{Shell \MakeLowercase{\textit{et al.}}: Bare Demo of IEEEtran.cls for IEEE Journals}

\maketitle

\begin{abstract}
Comprehensive semantic segmentation on renal pathological images is challenging due to the heterogeneous scales of the objects. For example, on a whole slide image (WSI), the cross-sectional areas of glomeruli can be 64 times larger than that of the peritubular capillaries, making it impractical to segment both objects on the same patch, at the same scale. To handle this scaling issue, prior studies have typically trained multiple segmentation networks in order to match the optimal pixel resolution of heterogeneous tissue types. This multi-network solution is resource-intensive and fails to model the spatial relationship between tissue types. In this paper, we propose the Omni-Seg network, a scale-aware dynamic neural network that achieves multi-object (six tissue types) and multi-scale (5$\times$ to 40$\times$ scale) pathological image segmentation via a single neural network. The contribution of this paper is three-fold: (1) a novel scale-aware controller is proposed to generalize the dynamic neural network from single-scale to multi-scale; (2) semi-supervised consistency regularization of pseudo-labels is introduced to model the inter-scale correlation of unannotated tissue types into a single end-to-end learning paradigm; and (3) superior scale-aware generalization is evidenced by directly applying a model trained on human kidney images to mouse kidney images, without retraining. By learning from ~150,000 human pathological image patches from six tissue types at three different resolutions, our approach achieved superior segmentation performance according to human visual assessment and evaluation of image-omics (i.e., spatial transcriptomics). The official implementation is available at \url{https://github.com/ddrrnn123/Omni-Seg}.

\end{abstract}


%
\IEEEpeerreviewmaketitle

\section{Introduction}
\IEEEPARstart {T}{he} process of digitizing glass slides using a whole slide image (WSI) scanner-known as  ``digital pathology" - has led to a paradigm shift in pathology~\cite{bandari2016renal}. Digital pathology not only liberates pathologists from local microscopes to remote monitors, but also provides an unprecedented opportunity for computer-assisted quantification~\cite{bengtsson2017computer, marti2021digital,gomes2021building}.
For example, the segmentation of multiple tissue structures on renal pathology provides disease-relative quantification by pathological morphology~\cite{wijkstrom2018morphological}, which is error-prone with variability by human visual examniation~\cite{zheng2021deep}. Many prior arts have developed pathological image segmentation approaches for pixel-level tissue characterization, especially with deep learning methods~\cite{kumar2017dataset,ding2020multi,ren2017computer,bel2018structure}. However, comprehensive semantic (multi-label) segmentation on renal histopathological images is challenging due to the heterogeneous scales of the objects. For example, the cross-sectional area of glomeruli can be 64 times larger than that of peritubular capillaries on a 2D WSI section~\cite{jensen1979influence}. Thus, human physiologists have to zoom in and out (e.g., between 40$\times$ and 5$\times$ magnifications) when visually examining a tissue in practice~\cite{li2020multiscale}. To handle this scaling issue, prior studies~\cite{jayapandian2021development,li2019u,hermsen2019deep} typically trained multiple segmentation networks that matched the optimal pixel resolution for heterogeneous tissue types. This multi-network solution is resource-intensive and its model fails to consider the spatial relationship between tissue types.

\begin{figure}[t]
\centering 
\includegraphics[width=0.9\linewidth]{{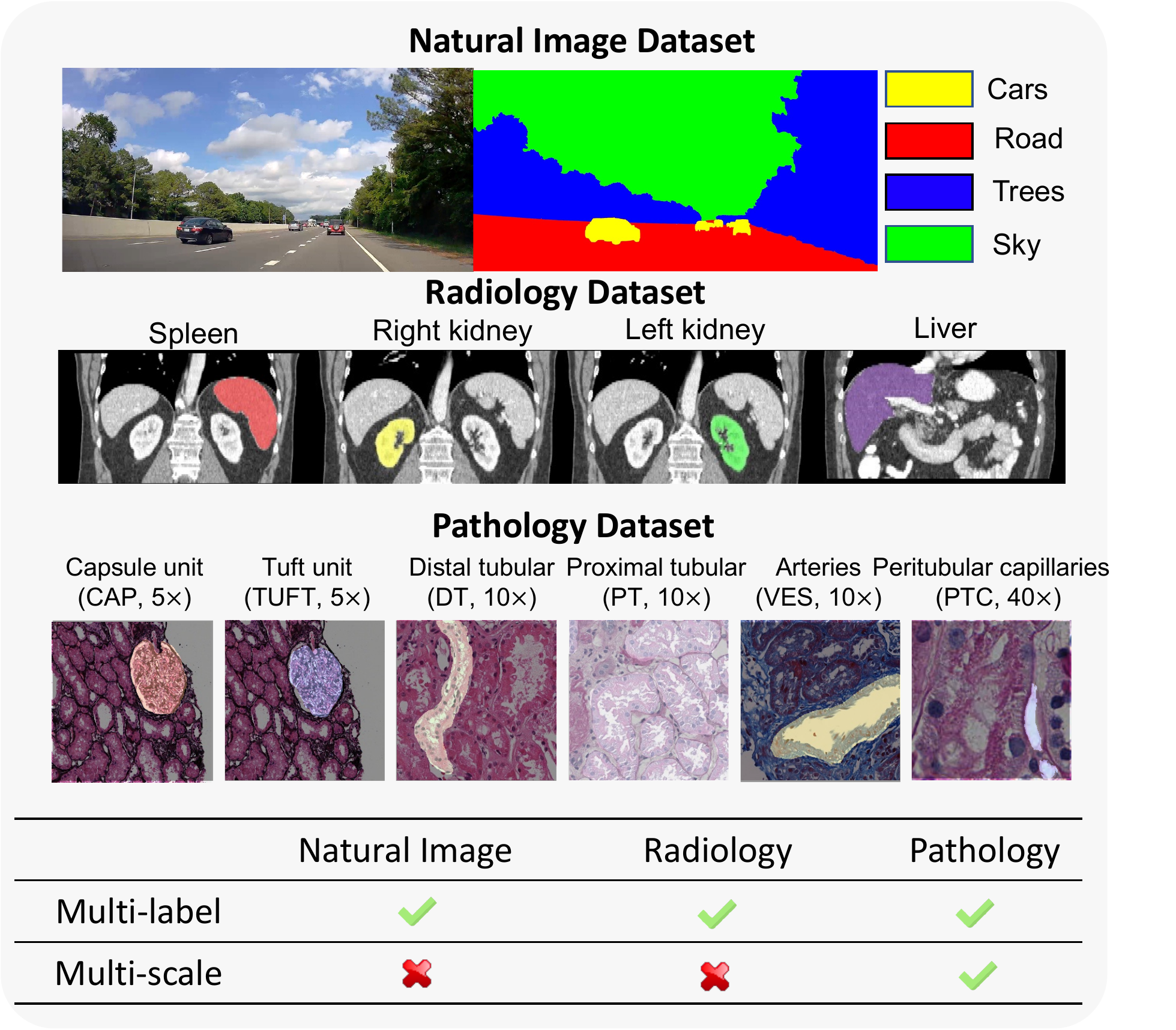}}
\caption{
\textbf{Challenge of multi-label multi-scale segmentation in renal pathology -- }
The semantic segmentation tasks are typically performed on a single scale for natural and radiological images. However, the multi-scale nature of the digitized pathological images (e.g., image pyramid in WSI) leads to a unique challenge of segmenting different tissue types at different optimal scales.} 
\label{Fig.problem} 
\end{figure}

\begin{figure*}[t]
\centering 
\includegraphics[width=0.9\linewidth]{{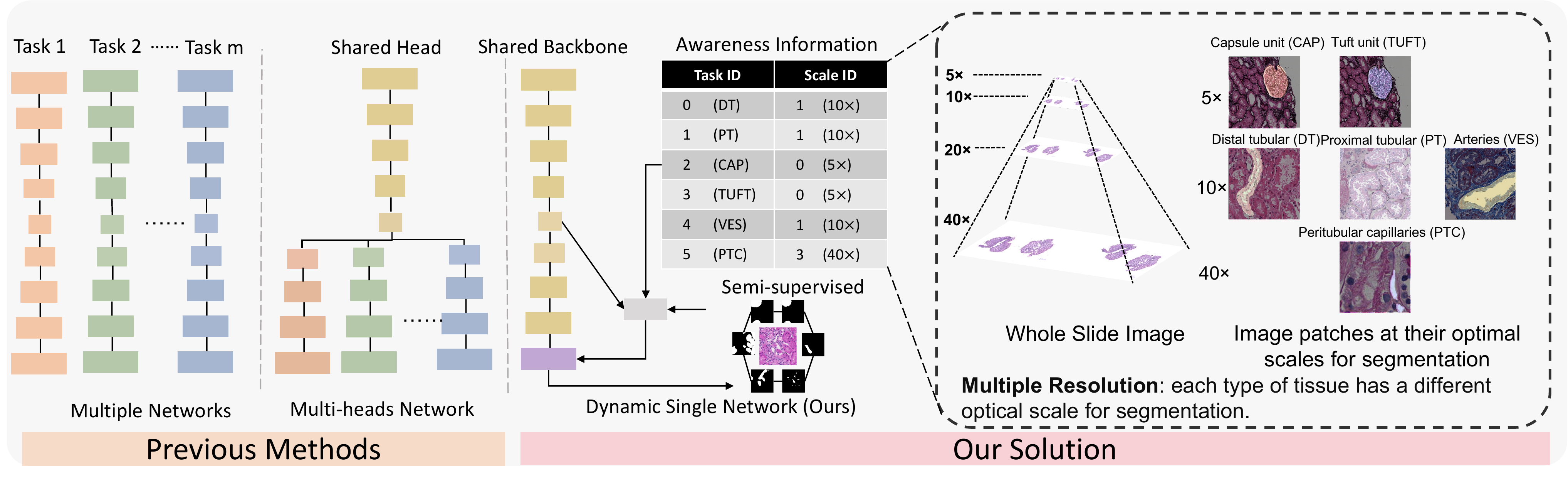}}
\caption{
\textbf{Dynamic neural networks design -- }
Previous work resolved multi-label multi-scale pathology segmentation through the utilization of multiple networks or multi-head networks (left panel). Advanced by recent dynamic neural networks, we propose a dynamic neural network method that aggregates both class-aware and scale-aware information into a single dynamic neural network. A semi-supervised learning strategy is further introduced to enhance the robustness of segmentation.
} 
\label{Fig.relativework} 
\end{figure*}

Recent advances in dynamic neural networks shed light on segmenting comprehensive tissue types via a single multi-label segmentation network~\cite{zhang2021dodnet,liu2022moddrop++,wu2022tgnet}. Dynamic neural networks generate the parameters of a neural network (e.g., the last convolutional layer) adaptively in the testing stage, achieving superior segmentation performance via a single network on various applications in natural and radiological image analysis. However, the multi-scale nature of the digitized pathological images (e.g., a WSI pyramid) leads to the unique challenge of adapting the Dynamic Neural Networks to pathology~\cite{barisoni2013digital}. For instance, Jayapandian et al.~\cite{jayapandian2021development} showed that the optimal resolution for segmenting glomerular units and tufts is 5$\times$, while the optimal resolution for segmenting the much smaller peritubular capillaries is 40$\times$.

 In this paper, we propose a single segmentation network, Omni-Seg, that performs multi-label multi-scale semantic segmentation on WSIs via a single dynamic neural network trained end-to-end. OmniSeg explicitly models the scale information as a scale-aware controller to, for the first time, make a single dynamic segmentation network aware of both scale information and tissue types in pathological image segmentation. The design is further generalized by introducing semi-supervised consistency regularization to model the spatial relationships between different tissue types even with different optimal segmentation scales. We evaluate the proposed method using the largest public multi-tissue segmentation benchmark in renal pathology, involving the glomerular tuft (TUFT), glomerular unit (CAP), proximal tubular (PT), distal tubular (DT), peritubular capillaries (PTC), and arteries (VES) with four different stains [Hematoxylin and Eosin (H\&E), Periodic-acid-Schiff (PAS), Silver (SIL), and Trichrome (TRI)] at three digital magnifications (5$\times$, 10$\times$, 40$\times$). 
 
 This work extended our conference paper \cite{deng2022omniseg} with new efforts as well as the contribution listed below: (1) a novel scale-aware controller is proposed to generalize the dynamic neural network from single-scale to multi-scale; (2) semi-supervised consistency regularization of pseudo-labels is introduced to model the inter-scale correlation of unannotated tissue types; and (3) superior scale-aware generalization of the proposed method is achieved by directly applying a model trained on human kidney images to mouse kidney images, without retraining. The code has been made publicly available at \url{https://github.com/ddrrnn123/Omni-Seg}.

\section{Related Works}
\subsection{Renal pathology segmentation}


With the recent advances in deep learning, Convolutional Neural Networks (CNNs) have become the \textit{de facto} standard method for image segmentation~\cite{feng2022artificial,hara2022evaluating}. Gadermayr et al.~\cite{gadermayr1708cnn} proposed two CNN cascades for histological segmentation with sparse tissue-of-interest. Gallego et al.~\cite{gallego2018glomerulus} implemented AlexNet for precise classification and detection using pixel-wise analysis. Bueno et al.~\cite{bueno2020glomerulosclerosis} introduced SegNet-VGG16 to detect glomerular structures through multi-class learning in order to achieve a high Dice Similarity Coefficient  (DSC). Lutnick et al.~\cite{lutnick2019integrated} implemented DeepLab v2 to detect sclerotic glomeruli and interstitial fibrosis and tubular atrophy region. Salvi et al.~\cite{salvi2021automated} designed mutliple residual U-Nets for glomerular and tubule quantification. Bouteldja et al.~\cite{bouteldja2021deep} developed a CNN for the automated multi-class segmentation of renal pathology for different mammalian species and different experimental disease models. Recently, instance segmentation approaches and Vision Transformers (ViTs) have been introduced to pathological image segmentation~\cite{johnson2019automatic,nguyen2021evaluating}. However, most of these approaches mainly focused on single tissue segmentation, such as glomerular segmentation with identification~\cite{gupta2018iterative,kannan2019segmentation, marechal2022automatic}. Moreover, there were several approaches are developed for disease-positive region segmentation~\cite{jing2022segmentation,lin2022adversarial}, rather than comprehensive structure understanding on renal pathology.

The conference version of Omni-Seg~\cite{deng2022omniseg}, utilizes a single residual U-Net as its backbone~\cite{ronneberger2015u,he2016deep} with a dynamic head design to achieve multi-class pathology segmentation. In this paper, we build upon our previous work by using a scale-aware vector to describe the scale-specific features and training the model with semi-supervised consistency regularization to understand spatial inferences between multiple tissue types at multiple scales, combining the information that is essential for pathological image segmentation.

\begin{figure*}[t]
\centering 
\includegraphics[width=0.9\textwidth]{{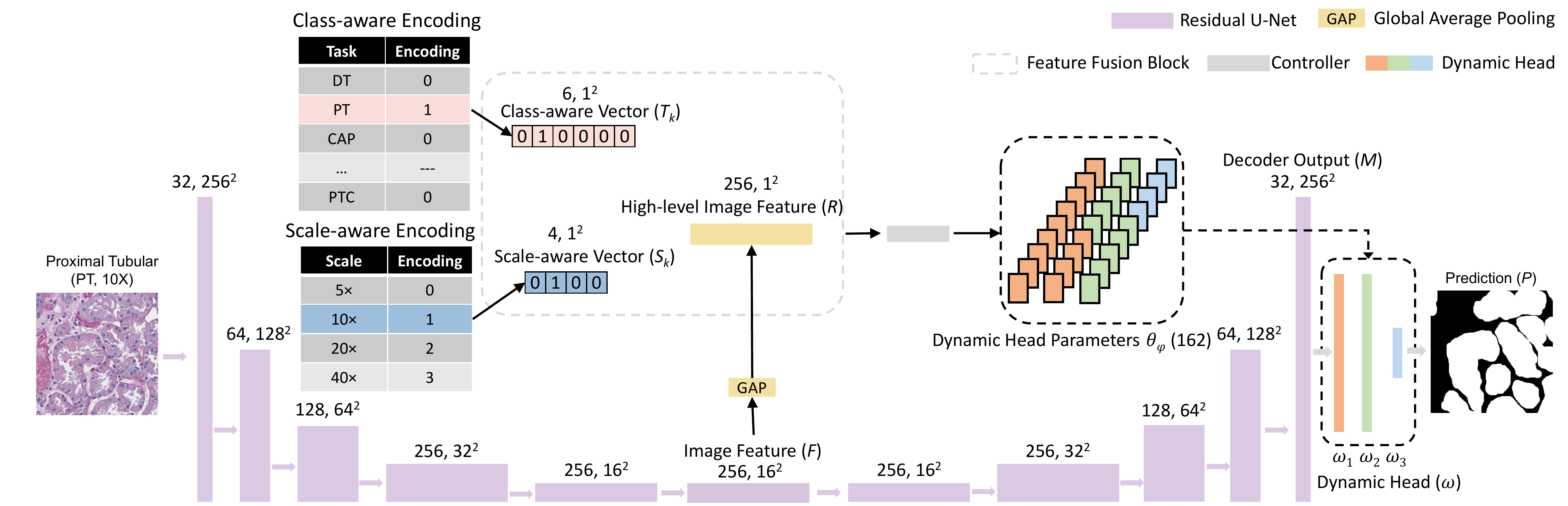}}
\caption{
\textbf{Omni-Seg Pipeline -- }
The proposed method consists of a residual U-Net backbone, a class-aware and a scale-aware controller, and a single dynamic segmentation head. A class-aware knowledge encoder and a scale-aware knowledge encoder are implemented for the multi-label and multi-scale pathological image segmentation. A feature-based fusion block is used to aggregate the features into the final dynamic head parameters.
} 
\label{Fig.Pipeline} 
\end{figure*}

\subsection{Multi-label medical image segmentation}
Deep learning-based segmentation algorithms have shown the capability of performing multi-label medical image segmentation~\cite{hermsen2019deep,bouteldja2021deep,jayapandian2021development}. Due to the issue of partial labeling, most approaches~\cite{jayapandian2021development,li2019u,hermsen2019deep} rely on an integration strategy to learn single segmentation from one network. This multi-network solution is resource intensive and suboptimal, without explicitly modeling the spatial relationship between tissue types. To address this issue, many methods have been proposed to investigate the partial annotation of a medical image dataset. Chen et al.~\cite{chen2019med3d} designed a class-shared encoder and class-specific decoders to learn a partially labeled dataset for eight tasks. Fang et al.~\cite{fang2020multi} proposed target-adaptive loss (TAL) to train the network by treating voxels with unknown labels as the background.


Our proposed method, Omni-Seg, was inspired by DoDNet~\cite{zhang2021dodnet}, which introduced the dynamic filter network to resolve multi-task learning in a partially labeled dataset. As shown in Fig.~\ref{Fig.relativework}, we generalized the multi-label DoDNet to a multi-label and multi-scale scenario. An online semi-supervised consistency regularization of pseudo-label learning extended the partially labelled dataset to the densely labelled dataset with non-overlap pseudo-labels. 


\subsection{Multi-scale medical image segmentation}
Unlike radiological images, pathological images contain multi-resolution images, called image pyramids, that allow different tissue types to be examined at their optimal magnifications or best resolutions~\cite{zhang2021dodnet}. However, modeling scale information for segmentation models is still challenging. Several deep learning-based approaches have been developed to aggregate scale-specific knowledge within the network architecture~\cite{chen2017rethinking,zhang2021pyramid,fu2018joint,gu2019net,sinha2020multi}. However, such technologies focus on feature aggregation from different scales and fail to learn scale-aware knowledge for heterogeneous tasks.

In our proposed network, we explicitly modeled and controlled pyramid scales (5$\times$, 10$\times$,20$\times$, 40$\times$) for a U-Net architecture by using a scale-aware controller joined with a class-aware controller by a feature fusion block. A scale-aware vector is proposed to encourage the network to learn distinctive features at different resolutions.

\section{Methods}
The overall framework of the proposed Omni-Seg method is presented in Fig.\ref{Fig.Pipeline}. The backbone structure is a residual U-Net, inspired by the existing multi-label segmentation network DoDNet~\cite{zhang2021dodnet} and Omni-Seg~\cite{deng2022omniseg} methods.


\subsection{Simultaneous multi-label multi-scale modeling}
Omni-Seg method was recently proposed to achieve multi-label segmentation using dynamic neural network design~\cite{deng2022omniseg}. However, such a method is not optimized for the multi-scale image pyramids in digital pathology. Moreover, the context information across different scales is not explicitly utilized in the learning process. To develop a digital pathology optimized dynamic segmentation method, the proposed Omni-Seg method generalize the model-aware encoding vectors to a multi-modal multi-scale fashion, with: (1) $m$-dimensional one-hot vector for class-aware encoding and (2)  $n$-dimension one-hot vector for scale-aware encoding, where $m$ is the number of tissue types, and $n$ is the number of magnifications for pathological images. The encoding calculation follows the following equation:

\begin{equation}
    T_k = 
    \begin{cases}
    1, \quad if \quad k = i \\
    0, \quad otherwise
    \end{cases}
    k = 1,2,...,m
\label{eq:class-aware}
\end{equation}

\begin{equation}
    S_p = 
    \begin{cases}
    1, \quad if \quad p = j \\
    0, \quad otherwise
    \end{cases}
    p = 1,2,...,n
\label{eq:scale-aware}
\end{equation}

\noindent where $T_k$ is a class-aware vector of $i$th tissue, and $S_p$ is a scale-aware vector in $p$th scale.

\subsection{Feature fusion block with dynamic head mapping}
\begin{figure}
\centering 
\includegraphics[width=0.8\linewidth]{{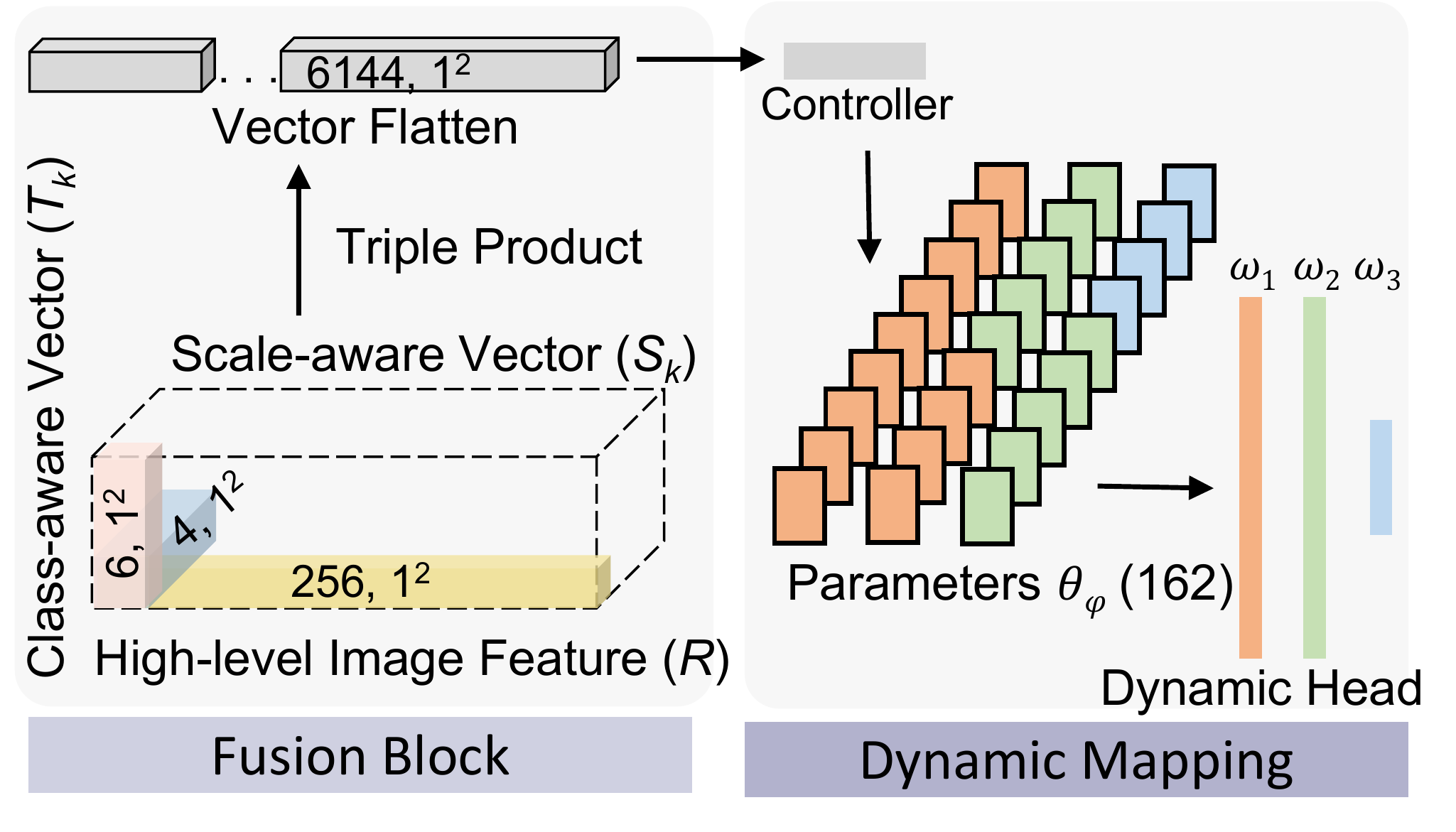}}
\caption{
\textbf{Feature Fusion Block-- }
A triple production is used to fuse three feature vectors from (1) the residual U-Net encoder, (2) multi-scale encoding features, and (3) multi-label encoding features. One CNN layer called the Controller is employed to receive the aggregated features from the triple product. It provides the required parameters for the dynamic head. The parameters used for the dynamic head are float values of output tensors from the Controller. $\omega_1$, $\omega_2$, and $\omega_3$ represent the parameters of the convolutional layers in the dynamic head (Eq.\ref{eq:dynamic-fusion}).
} 
\label{Fig.Fusion} 
\end{figure}

To provide the multi-class and multi-scale information to the embedded features, we combine two vectors into the low dimensional feature embedding at the bottom of the residual U-Net architecture. The image feature $F$ is summarized by a Global Average Pooling (GAP) and receives a feature vector in the shape $\mathbb{R}^{N\times256\times1\times1}$, where $N$ is batch-size. The one-hot-label class-aware vector $T_k$ ($\mathbb{R}^{N\times6}$) and the scale-aware vector $S_p$ ($\mathbb{R}^{N\times4}$) are reformed to $\mathbb{R}^{N\times6\times1\times1}$ and $\mathbb{R}^{N\times64\times1\times1}$, respectively, to match the dimensions with the image features for the next fusion step. Different from the conference version of Omni-Seg~\cite{deng2022omniseg} which directly concatenates the feature vectors, a triple outer product is implemented to combine three vectors into a one-dimensional vector by a flatten function, following a single 2D convolutional layer controller, $\varphi$, as a feature fusion block to refine the fusion vector as the final controller for the dynamic head mapping:

\begin{equation}
    \omega = \varphi(GAP(F)||T||S;\Theta_\varphi)
\label{eq:fusion-equation}
\end{equation}

\noindent where $GAP(F)$, $T$, and $S$ are combined by the fusion operation, $||$, and $\Theta_\varphi$ is the number of parameters in the dynamic head. The feature-based fusion implementation is shown in Fig.~\ref{Fig.Fusion}.

Inspired by~\cite{zhang2021dodnet}, a binary segmentation network is employed to achieve multi-label segmentation via a dynamic filter. From the multi-label multi-scale modeling above, we derive joint low-dimensional image feature vectors, class-aware vectors, and scale-aware vectors at an optimal segmentation magnification. The information is then mapped to control a light-weight dynamic head, specifying (1) the target tissue type and (2) the corresponding pyramid scale.

The dynamic head concludes with three layers. The first two have eight channels, while the last layer has two channels. We directly map parameters from the fusion-based feature controller to the kernels in the 162-parameter dynamic head to achieve precise segmentation from multi-modal features. Therefore, the filtering process can be expressed by Eq.\ref{eq:dynamic-fusion}

\begin{equation}
    P = ((((M * \omega_1) * \omega_2) * \omega_3)
\label{eq:dynamic-fusion}
\end{equation}

\noindent where $*$ is convolution, $P\in\mathbb{R}^{N \times 2\times W \times H}$ is the final prediction, and $N$, $W$, and $H$ correspond to the batch-size, width, and height of the dataset, respectively.

\begin{figure}
\centering 
\includegraphics[width=1\linewidth]{{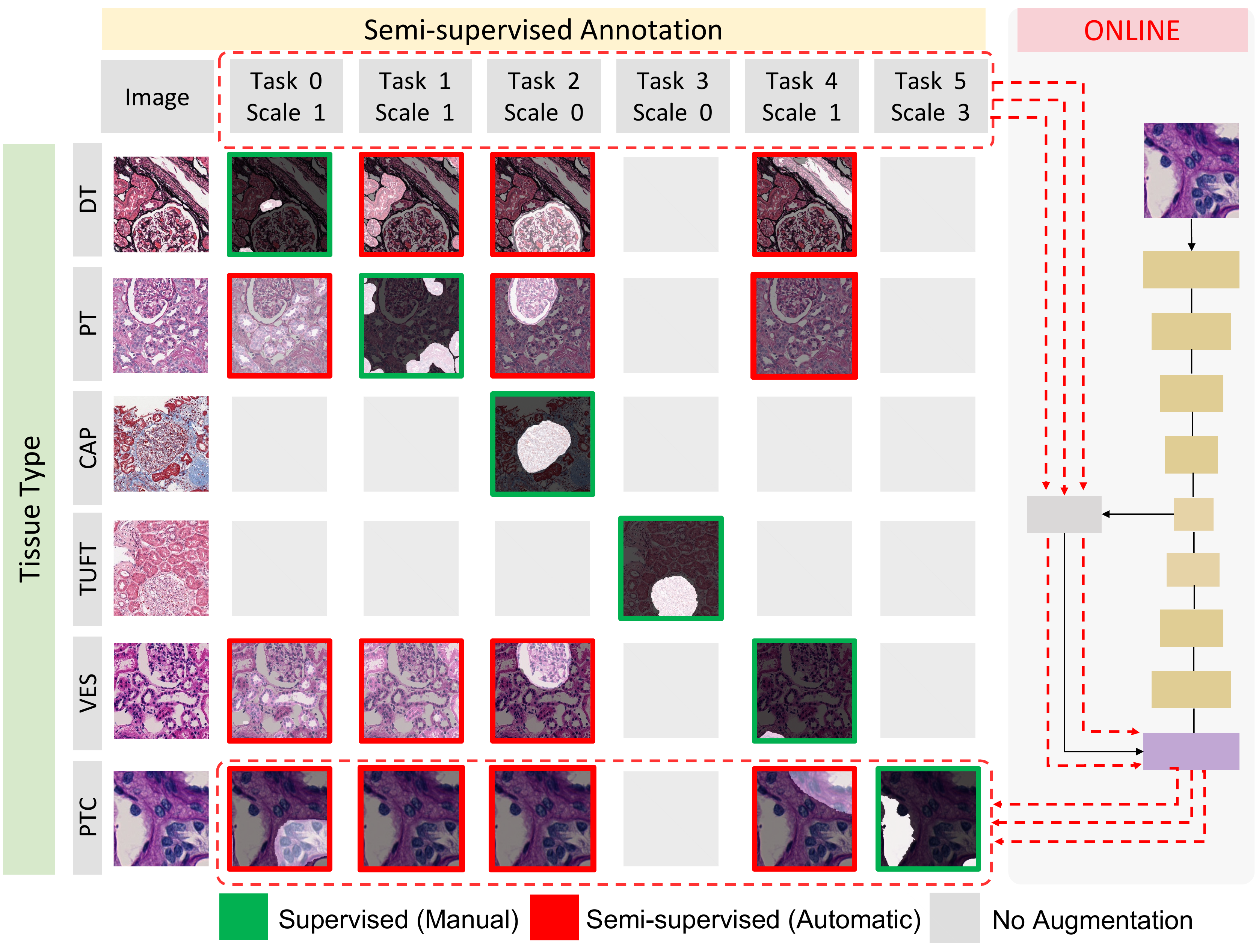}}
\caption{
\textbf{Pseudo label dataset -- }
Pseudo-labels are used in the semi-supervised learning strategy so as to impute the missing class-aware vectors and scale-aware vectors (red boxes) from the labeled training data (green boxes). Pseudo labels for segmenting CAPs and TUFTs are not included since those tissue types are relatively easy to segment among all tissue types.
} 
\label{Fig.Pseudolabel} 
\end{figure}

\begin{figure}
\centering 
\includegraphics[width=0.6\linewidth]{{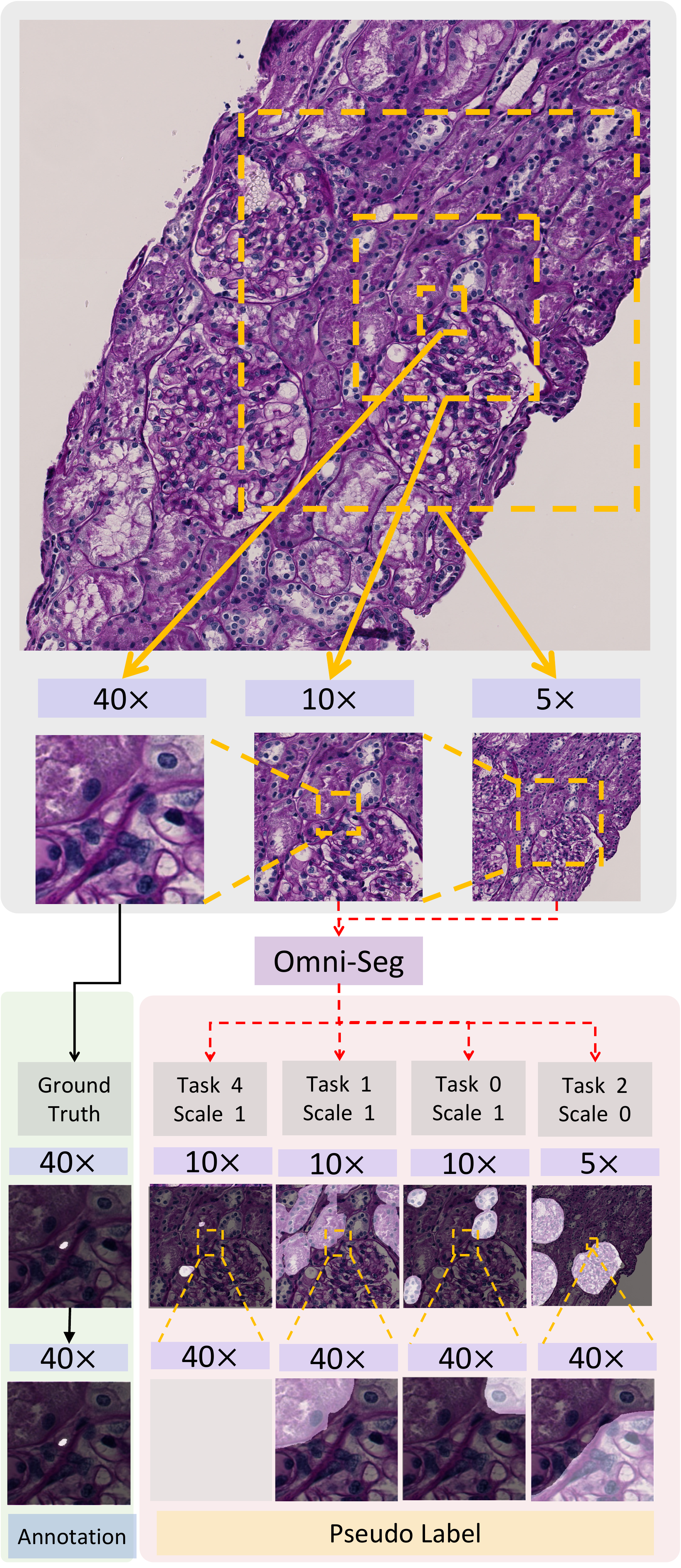}}
\caption{
\textbf{Matching Selection -- }
The semi-supervised pseudo-labels are introduced to the Omni-Seg to utilize the inter-tissue type correspondence. Briefly, the patches from the supervised training data are extracted with the pseudo-labels on the corresponding locations. The Table~\ref{table:differentdesigns} demonstrates that this strategy achieved better performance than using all pseudo labels with grid patches, which may lose the feature correlation from the existing ground truth while adding more interferences. 
} 
\label{Fig.MatchingPatch} 
\end{figure}

\subsection{Semi-supervised consistency regularization of pseudo Label learning}

An online semi-supervised pseudo-label learning strategy is proposed to generate the ``densely labeled” dataset for the learning of spatial correlation. The original large images at 40$\times$ magnification are tiled into small patches with multiple sizes and downsampled to a size of 256$\times$256 pixel resolution to rescale their magnifications to the optimal resolutions, respectively. At each scale, the patches are segmented for multiple tissues at their optimal segmentation magnification by using different class-aware vectors and scale-aware vectors. Then, the patches are aggregated back into the original 40$\times$ physical space according to their original location and are then rescaled. There are two strategies for collecting the ``densely labeled" dataset with pseudo-labels at the patch level. The first one is tiling the large images into different scales with a 256$\times$256 pixel resolution, while the second one uses a similarity score to locate the patches in the supervised training data, matching and cropping the consistent area pseudo-labels. The matching selection is shown in Fig.~\ref{Fig.MatchingPatch}. As a result of the ablation study in Table~\ref{table:differentdesigns}, the matching selection attained a better performance with a better understanding of spatial relationships between supervised labels and pseudo-labels. Fig.~\ref{Fig.Pseudolabel} demonstrates the online ``densely labeled" dataset with extended pseudo-labels. The pseudo-labels expand the dimensional correspondences for multiple tissues at multiple resolutions. Inspired, ~\cite{chen2021exploring}, a semi-supervised constraint is introduced to enforce the similar embedding of two augmentations upon the same images.

\section{Data and Experimental Design}
\subsection{Data}
1,751 regions of interest (ROIs) images were captured from 459 WSIs, obtained from 125 patients with Minimal Change Diseases. The images were manually segmented for six structurally normal pathological primitives~\cite{jayapandian2021development}, using the digital renal biopsies from the NEPTUNE study~\cite{barisoni2013digital}. All of the images had a resolution of 3000$\times$3000 pixels at a 40$\times$ magnification (0.25 $\mu m$ pixel resolution), including TUFT, CAP, PT, DT, PTC, and VES in H\&E, PAS, SIL, and TRI stain. Four stain methods were regarded as color augmentations in each type of tissue. The study exempt from IRB approval by Vanderbilt University Medical Center IRB board. We followed~\cite{jayapandian2021development} to randomly crop and resized them into 256$\times$256 pixels resolution. We kept the same splits as the original release in ~\cite{jayapandian2021development}, where the training, validation, and testing samples were separated with a 6:1:3 ratio. The splits were performed at the patient level to avoid data contamination.

\subsection{Experimental Design}

The entire training process was divided into two parts. In the first 50 epochs, only a supervised learning strategy was employed to minimize the binary dice loss and cross-entropy loss. Then, both supervised and semi-supervised learning were executed to explore the spatial correlation between multiple tissues with multiple resolutions. For the semi-supervised learning, four supervised training patches originally from the full size 40$\times$ original image were randomly selected to generate pseudo labels for DT, PT, CAP, TUFT, and VES, while 16 patches were randomly selected for PTC. Beyond the binary dice loss and cross-entropy loss, KL Divergence loss and Mean-Squared-Error loss were used as extra semi-supervised constraints with different image augmentations. The SGD was used as the optimizer in both supervised and semi-supervised learning.

\begin{figure*}
\centering 
\includegraphics[width=0.9\linewidth]{{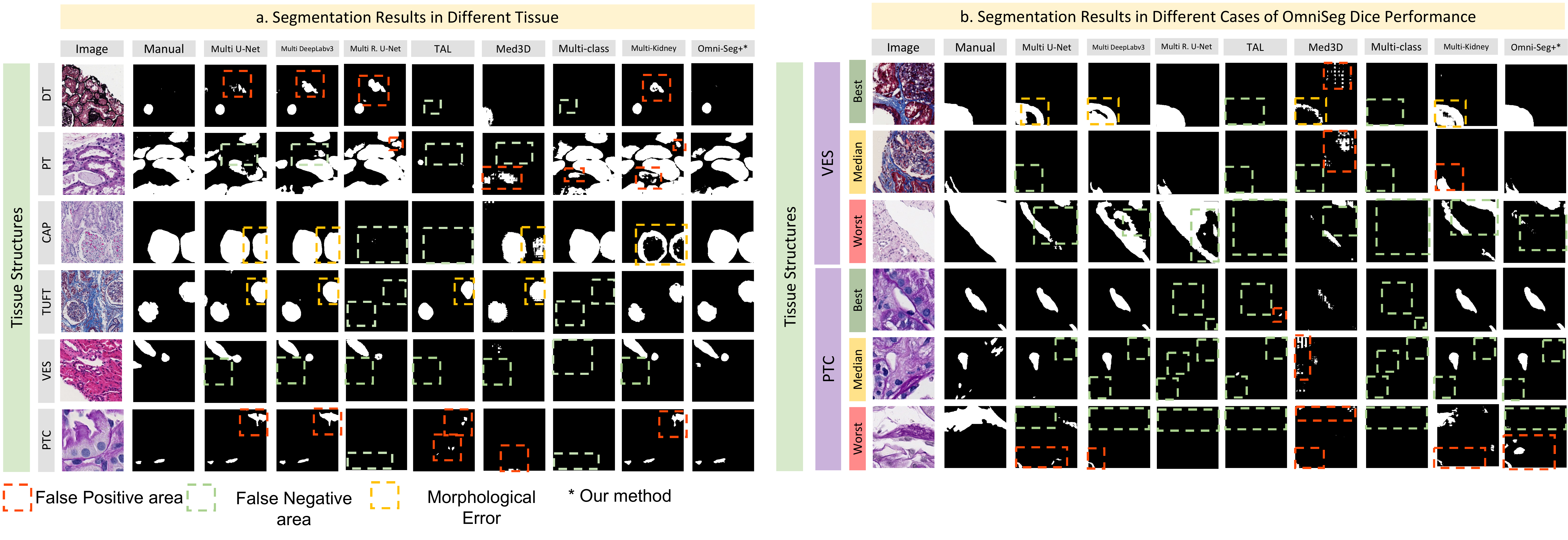}}
\caption{
\textbf{Internal Validation Qualitative Results -- }This figure shows the qualitative results of different approaches. The red, green, and yellow bounding boxes present the false positive, false negative, and morphological errors in the predicted masks, respectively. The cases with best, median, and worst Dice scores are provided for a qualitative comparison. }

\label{Fig.HumanResults} 
\end{figure*}

\section{Results}
We compared the proposed Omni-Seg network to baseline models, including (1) multiple individual U-Net models (U-Nets)~\cite{jayapandian2021development}, (2) multiple individual DeepLabv3 models (DeepLabv3s)~\cite{lutnick2019integrated},  and (3) multiple individual Residual U-Nets models~\cite{salvi2021automated} for renal pathology quantification. We also compared the proposed network to (4) a multi-head model with target adaptive loss (TAL) for multi-class segmentation~\cite{fang2020multi}, (5) a multi-head 3D model (Med3D) for multiple partially labeled datasets~\cite{chen2019med3d}, (6) a multi-class segmentation model for partially labeled datasets~\cite{gonzalez2018multi}, and (7) a multi-class kidney pathology model~\cite{bouteldja2021deep}. All of the parameter settings are followed by original paper.

\begin{table*}
\centering
\caption{
Performance of different methods on the human dataset (internal validation).}


\begin{tabular}{l|ccccccccc}
\toprule
\multirow{2}{0.8in}{Method} & \multicolumn{3}{c}{DT (10$\times$)} & \multicolumn{3}{c}{PT (10$\times$)} & \multicolumn{3}{c}{CAP (5$\times$)}\\
\cmidrule(lr){2-4}
\cmidrule(lr){5-7}
\cmidrule(lr){8-10}
 & Dice$\uparrow$ & HD$\downarrow$ & MSD$\downarrow$ & Dice$\uparrow$ & HD$\downarrow$ & MSD	$\downarrow$ & Dice$\uparrow$ & HD$\downarrow$ & MSD$\downarrow$ \\
\midrule
U-Nets~\cite{jayapandian2021development} & 78.51 & 107.63 & 36.05 & 88.25 & 63.84 & 8.53 & 95.42 & 54.38 & 9.34 \\
DeepLabV3s~\cite{lutnick2019integrated} & 77.92 & 107.61 & 35.45 & 88.49 & 60.24 & 9.05 & 95.78 & \textbf{50.42} & 9.76 \\
Residual U-Nets~\cite{salvi2021automated} & 78.60 & 107.03 & 31.64 
&88.71 & 62.20 & 8.41 & 57.85 & 325.26 & 201.74  \\
\midrule
TAL~\cite{fang2020multi} & 47.76 & 280.44 & 198.07 & 48.49 & 179.41 & 84.91 & 51.82 & 402.76 & 272.76 \\
Med3D~\cite{chen2019med3d} & 47.73 & 194.55 & 110.09 & 35.80 & 217.41 & 109.51 & 89.49 & 89.76 & 19.92 \\

Multi-class~\cite{gonzalez2018multi} & 47.76	& 280.44&	198.07 & 88.36 &	64.84&	8.85 & 95.93&	85.64&	11.1 \\

Multi-kidney~\cite{bouteldja2021deep} & 80.52	& 102.21&	24.49 & 89.23 &	61.98&	8.07 & 81.47 &	104.04&	19.80 \\

\midrule
Omni-Seg (Ours) & \textbf{81.11} & \textbf{97.99} & \textbf{22.54} &\textbf{89.86} & \textbf{56.85} & \textbf{6.78} & \textbf{96.70} & 51.81 & \textbf{7.37} \\
\bottomrule

\end{tabular}

\begin{tabular}{l|cccccccccccc}
\toprule
\multirow{2}{0.8in}{Method} & \multicolumn{3}{c}{TUFT (5$\times$)} & \multicolumn{3}{c}{VES (10$\times$)} & \multicolumn{3}{c}{PTC (40$\times$)} & \multicolumn{3}{c}{Average}\\
\cmidrule(lr){2-4}
\cmidrule(lr){5-7}
\cmidrule(lr){8-10}
\cmidrule(lr){11-13}
 & Dice$\uparrow$ & HD$\downarrow$ & MSD$\downarrow$ & Dice$\uparrow$ & HD$\downarrow$ & MSD$\downarrow$ & Dice$\uparrow$ & HD$\downarrow$ & MSD$\downarrow$ & Dice$\uparrow$ & HD$\downarrow$ & MSD$\downarrow$\\
\midrule

U-Nets~\cite{jayapandian2021development} & 96.05 &	63.16&	9.72&	77.66&	101.59&	54.30&	72.73&	31.80	&13.53 &  84.77 & 70.4 & 21.91 \\
DeepLabV3s~\cite{lutnick2019integrated} & 96.45&	51.34&	7.16&	81.08&	84.31&	44.69&	72.69&	30.75&	14.27 & 85.40 & 64.11 & 20.06\\
Residual U-Nets~\cite{salvi2021automated} & 54.59 & 367.68 & 247.08 & 76.71 & 95.37 & 46.43 & 49.22 & 64.71 & 49.59 & 67.61 & 170.38 & 96.98 \\
\midrule

TAL~\cite{fang2020multi} & 76.95 &	137.18 &	65.58 &	47.67 &	244.11 & 191.25 &	49.37 &	52.15 &	35.79 & 53.67 & 216.01 & 141.39\\

Med3D~\cite{chen2019med3d} & 92.80 &	80.84 &	14.80 &	58.46 &	150.71 &	78.77 &	49.78 &	44.53 &27.24 & 62.34 & 129.63 & 60.06\\

Multi-class~\cite{gonzalez2018multi} & 46.63&	486.30&	359.04 &	47.67&	244.12&	191.20 &	49.28	&64.38	&49.41& 62.69&	204.28&	136.27\\

Multi-kidney~\cite{bouteldja2021deep}& 82.24 &	84.00&	21.32 & 83.74&	87.46&	25.19 &	75.97	&28.76	& 9.07& 82.20&	78.08 &	17.99\\

\midrule

Omni-Seg (Ours) & \textbf{96.66}&    \textbf{39.93}&	\textbf{5.71}&	\textbf{85.02}&	\textbf{74.83}&	\textbf{22.29}&	\textbf{77.19}&	\textbf{25.61}&	\textbf{7.87}& \textbf{87.76} & \textbf{57.84} & \textbf{12.09} \\

\bottomrule
\end{tabular}
\label{table:humandataset}
\end{table*}

\begin{table}
\begin{center}
\caption{
Performance of different methods on different resolution of murine dataset (external validation).}

\begin{tabular}{l|cc}
\toprule
Method  & PT (10$\times$)  & CAP (20$\times$*)\\
\midrule
U-Nets~\cite{jayapandian2021development}  & 71.05 &  89.30\\
DeepLabV3s~\cite{lutnick2019integrated} & 73.15 & 89.80\\
Residual U-Nets~\cite{salvi2021automated} & 74.04 & 35.42\\
\midrule
TAL~\cite{fang2020multi} & 4.18  & nan \\
Med3D~\cite{chen2019med3d} & -5.16 & 85.89\\
Multi-class~\cite{gonzalez2018multi} & 61.25  & 87.39 \\
Multi-kidney~\cite{bouteldja2021deep} &  74.30 & 69.11\\
\midrule
Omni-Seg (Ours) & \textbf{75.25}  & \textbf{91.73} \\
\bottomrule
\end{tabular}
\label{table:differentresolution} 
\end{center}
\text{*The size of CAP on mouse kidney dataset at 20$\times$ is consistent with that on} \\
\text{human kidney dataset at 5$\times$ to match the size of human glomeruli.}

\end{table}

\begin{table}
\caption{Ablation study of Omni-Seg on the murine dataset.
}
\begin{center}
\begin{tabular}{lcclcc}
\hline
SC & MS & CR & PT & CAP\\
\hline
     & &  &	57.73 & 87.14 \\
     & & \checkmark & 64.32 & 87.15 \\
     &\checkmark &\checkmark &	66.12 & 90.04 \\
\hline
    \checkmark & &  &  72.80 & 89.64 \\
    \checkmark & & \checkmark & 69.80 & 83.52 \\
    \checkmark & \checkmark & \checkmark& 	\textbf{75.25} & \textbf{91.73} \\
\hline
\end{tabular}
\end{center}
\text{*SC is Scale-aware Controller}\\
\text{*MS is Matching Selection}\\
\text{*CR is Consistency Regularization}\\
\label{table:differentdesigns} 
\end{table}

\subsection{Internal validation}
Table~\ref{table:humandataset} and Fig.~\ref{Fig.HumanResults} show the results on the publicly available dataset~\cite{jayapandian2021development}. The distance metrics are in units of Micron. In Table~\ref{table:humandataset}, Omni-Seg achieved the better performance in most metrics. In Fig.~\ref{Fig.HumanResults}, Omni-Seg achieved better qualitative results with less false-positive, false-negative, and morphological errors among the best, the median, and the worst Dice cases. The Dice similarity coefficient (Dice: \%, the higher, the better), Hausdorff distance (HD, Micron unit: the lower, the better), and Mean Surface Distance (MSD, Micron unit, the lower, the better) were used as performance metrics for evaluating the quantitative performance.



\begin{figure}
\centering 
\includegraphics[width=1.0\linewidth]{{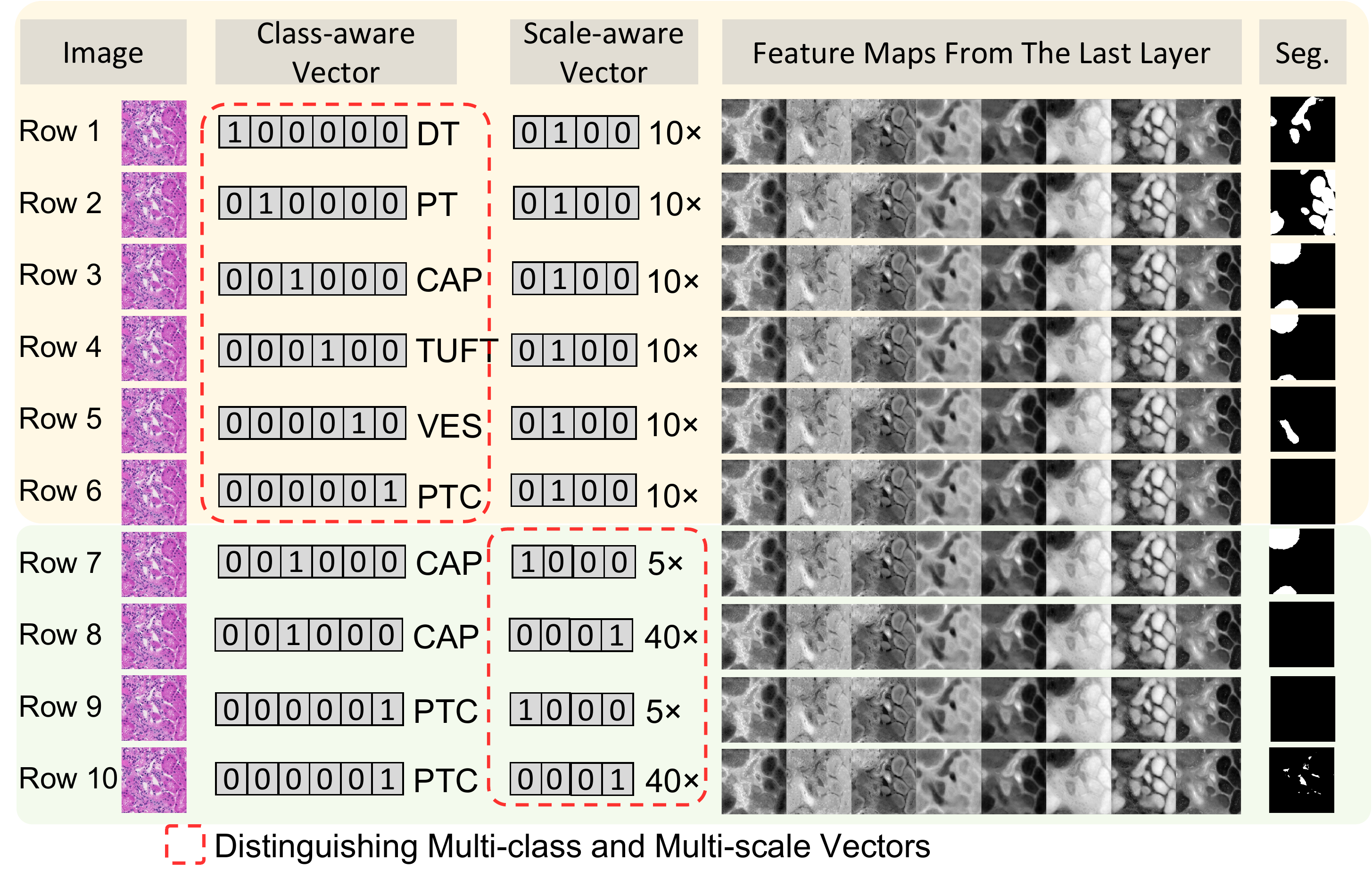}}
\caption{
\textbf{Intermediate representation -- }
This figure shows the unique advantage of the dynamic neural network design. Specifically, the single set of feature maps are shared by all organs and scales, while different segmentation outcomes are achieved with the multi-label multi-scale controllers.
} 
\label{Fig.Featuremap} 
\end{figure}

Fig.~\ref{Fig.Featuremap} illustrates the functionality of the multi-class and multi-scale dynamic design in Omni-Seg, with both intermediate representations and final segmentation masks. First, the shared feature maps are identical before applying the class-aware and scale-aware dynamic control. Then, different segmentation results are achieved for different tissue types (Row 1 to 6) and different scales (Row 7 to 10), from a single deep neural network. 

\begin{figure*}
\centering 
\includegraphics[width=0.75\linewidth]{{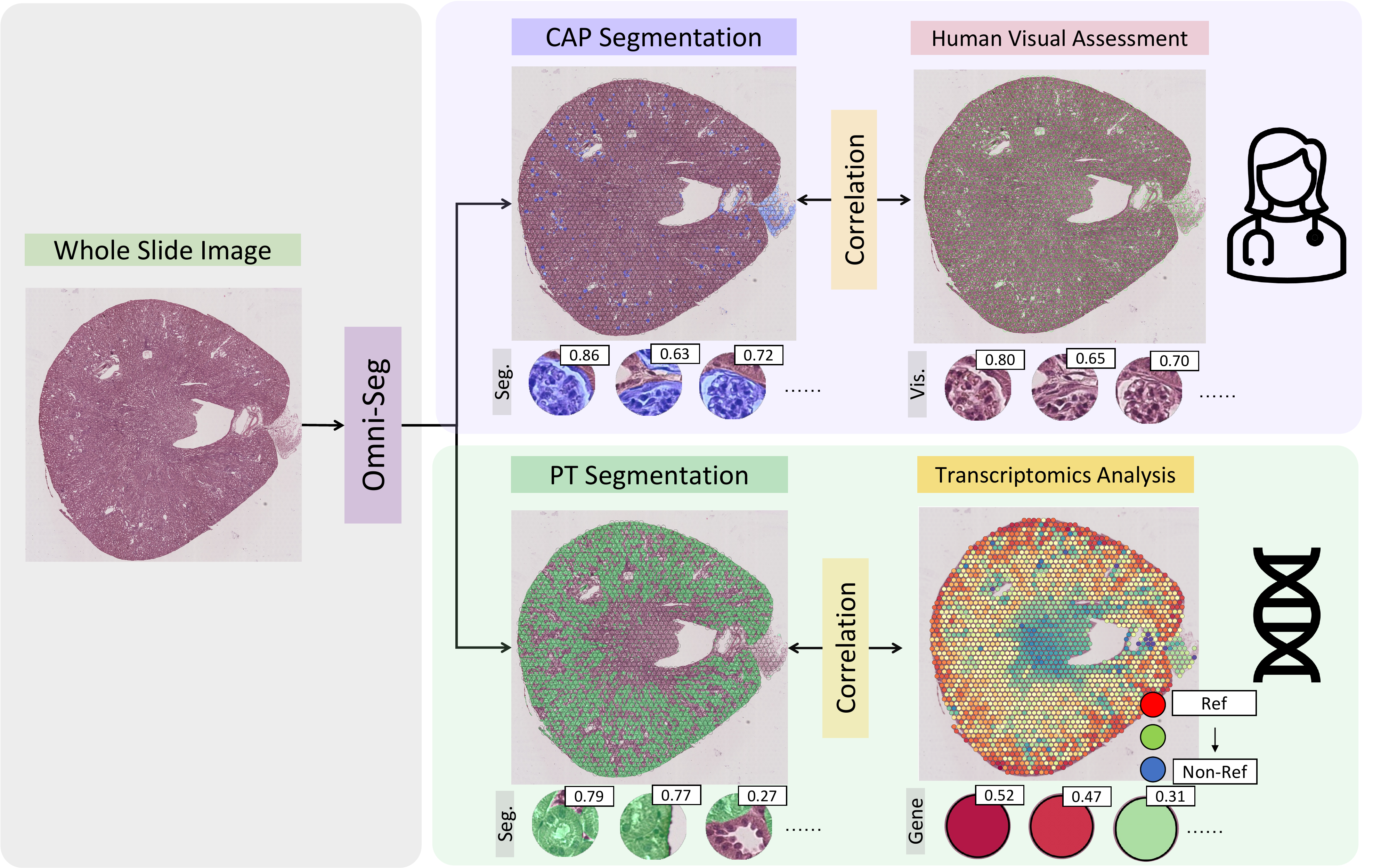}}
\caption{
\textbf{Evaluate the external segmentation performance using manual quantification and spatial transcripts -- }
The segmentation performance of the Omni-Seg on an external mouse WSI is presented. The percentage of the CAP tissue in each spot is compared with the visual estimation from pathologists. The percentage of the PT tissue in each spot is compared with the spatial transcriptomics results since it is difficult to perform such an estimation by human pathologists.} 
\label{Fig.Transcriptomics} 
\end{figure*}

\subsection{External Validation}
To validate our proposed method on another application, Omni-Seg was evaluated by directly applying the model trained on a human kidney dataset to a murine kidney dataset (without retraining). 

\subsubsection{Data}
Four murine kidneys were used as the external validation, with both H$\&$E WSIs (20$\times$) and 10$\times$ Visium spatial transcriptomics acquisition. All animal procedures were approved by the Institutional Animal Care and the Use Committee at Vanderbilt University Medical Center.


\subsubsection{Approach}
We applied different segmentation approaches (as shown in Table~\ref{table:differentresolution}) to the whole kidney WSI. We extracted the patches with 55 $\mu$m diameter (circle shaped spots) according to the 10$\times$ Visium spatial transcriptomics protocol~\cite{maynard2021transcriptome}. Then, we compared the proportions of the targeting tissue types in each spot with human labels and genetic labels (Fig.~\ref{Fig.Transcriptomics}). 

\textbf{CAP percentages in spots}. One pathologist was asked to label the percentage of CAP area in each spot, rather than performing resource-intensive pixel-level annotation. Then, such percentage can be automatically achieved from different segmentation methods. A Pearson correlation score was computed between the manual labels and automatic estimations, as shown in Table~\ref{table:differentresolution}. 

\textbf{PT percentages in spots}. It was difficult to replicate the above evaluation for PT since to visually differentiate PT from DT is challenging even for human pathologists. Fortunately, spatial transcriptomics analytics were able to offer the percentile of PT specific cell counts with in each spot. We believe this was the most unbiased approximation that was available to evaluate the PT segmentation. Briefly, the transcriptomics sequencing data were demultiplexed by “mkfastq” module in SpaceRanger~\cite{amezquita2020orchestrating}. fastQC~\cite{simons2010quality} were used for Quality control. R package Seurat~\cite{hao2021integrated} was used for data processing, while the spacexr~\cite{cable2022robust} software was employed to obtain the PT cell percentages via cell deconvolution. We compare such percentages with the ones from different automatic segmentation approaches, as shown in Table~\ref{table:differentresolution}.


\subsubsection{Experimental Details} 
PT and CAP were extracted with the diameter of the spots is 55 $\mu$m, which is 110 pixels on 20$\times$ digital WSIs, following the standard 10$\times$ Visium spatial transcriptomics protocol~\cite{maynard2021transcriptome}. 

\subsubsection{Results}
Table~\ref{table:differentresolution} shows the Pearson Correlation scores of CAP and PT percentages with human and spatial transcriptomics labels. Three digital magnifications (5$\times$, 10$\times$, 20$\times$) are generated by downsampling the 20$\times$ WSIs for a more comprehensive assessment. As a result, Omni-Seg achieved superior performance (in red) for most evaluations. The correlation metric of TAL for the capsule glomerular tissue is nan because of zero predictions for all patches.

\subsection{Ablation Studies}

\begin{figure}
\centering 
\includegraphics[width=0.75\linewidth]{{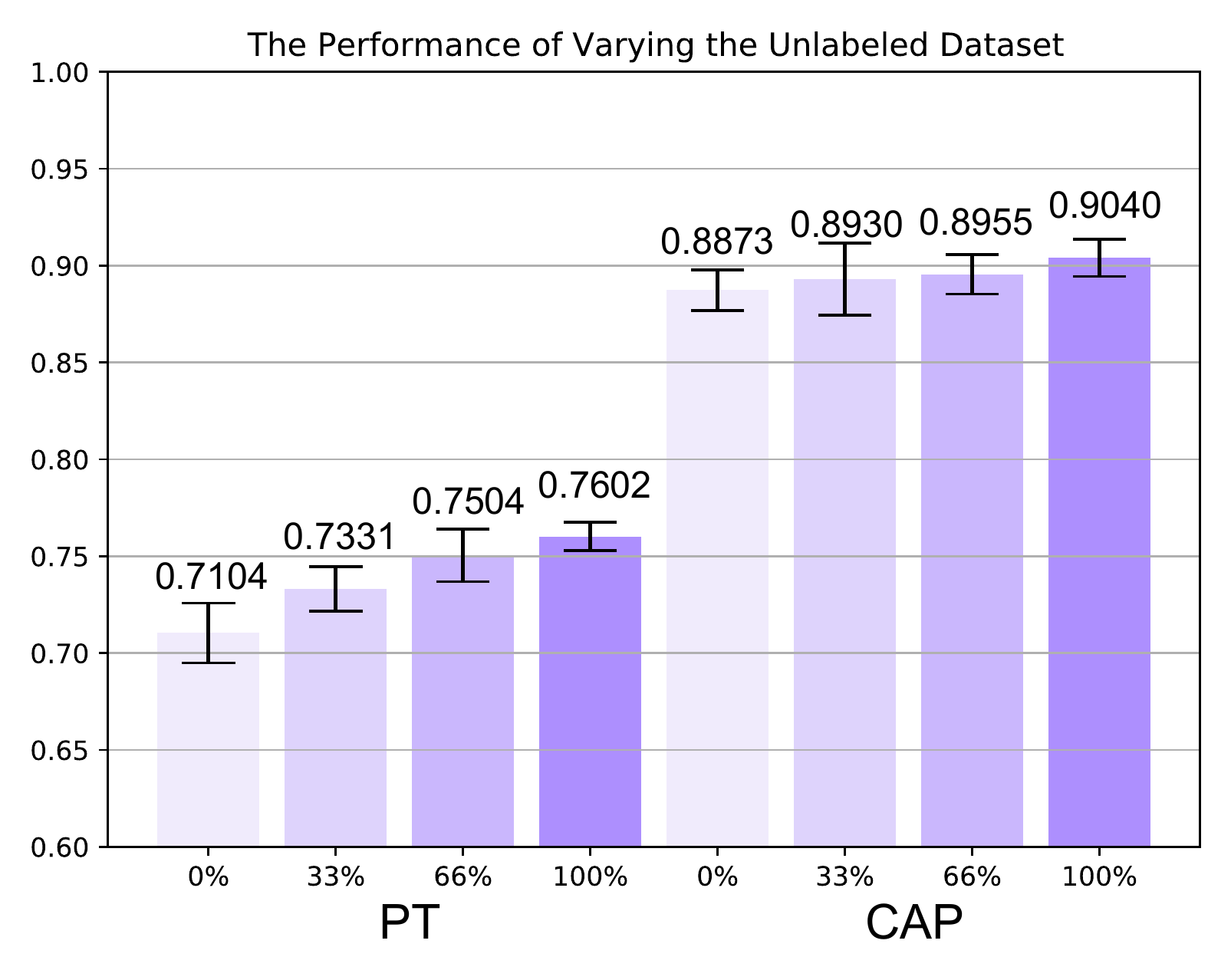}}
\caption{
\textbf{Evaluate the performance of varying semi-supervised dataset -- }
This figure shows the error bars of the results on varying sizes of the semi-supervised dataset with 5 repeated runs. The proposed semi-supervised consistency regularization of pseudo-label learning is evaluated by varying the unlabeled data set. The data split of 33$\%$ dataset is part of 66$\%$ dataset.} 
\label{Fig.Semidataset} 
\end{figure}

Table~\ref{table:differentdesigns} indicates the performance of the different model designs of Omni-Seg on the external validation dataset. The Omni-Seg approach with a Scale-aware Controller (SC), Matching Selection (MS), and Consistency Regularization (CR) achieved superior performance. We also evaluated our semi-supervised consistency regularization of pseudo-label learning by varying the unlabeled data set (Fig.~\ref{Fig.Semidataset}). The data split of 33$\%$ dataset is part of 66$\%$ dataset. To eliminate the unbalanced performance among different segmentation tasks, the model was repetitively trained for five times on each size of the dataset and get the mean values and standard deviation values of evaluation metrics. In general, the segmentation performance is monotonically increasing and more stable on each tissue type when enlarging the dataset. The model yields the comparable performance of using 66$\%$ of the available pseudo-label data, compared with the scenarios of using 100$\%$ of the cohort.

\section{Discussion}
In this study, we propose a novel single dynamic segmentation network with scale information for histopathology images. With the consistency regularization of multi-tissues and multi-scales on a consistent area of supervised training data, the proposed model can observe and extend the spatial relationship and the scale consistency from originally partially annotated multi-scale pathological images.

Table~\ref{table:humandataset} demonstrates that the proposed single network design can enhance 3\% of the overall DSC of segmentation by aggregating multi-class and multi-scale knowledge in a single backbone. Moreover, when applying both methods onto another independent datasest with different tissue scales, the Omni-Seg achieves overall superior performance compared with other bench marks (Table~\ref{table:differentresolution}).

There are several limitations and potential future improvements for our study. In the current version of the network, each region of the WSIs needs to be resized to the optimal resolution since all the tissues are segmented in different resolutions as a means of binary segmentation. Thus, it is a time consuming process to aggregate the tissue-wise segmentation results into the final multi-label segmentation masks, which increases the computational times during the testing stage.


The network provides morphological quantification for multiple tissues that can efficiently assist to the topography of gene expression in transcriptomics analysis for future genomics examinations. Meanwhile, the current single network with a class-aware vector and scale-aware vector can be simply applied to the additional dataset by fine-tuning the specific tissue types at different scales. Further work is needed to evaluate the proposed method's applicability to types of digital pathology datasets other than the ones explored here.

\section{Conclusion}
In this paper, we propose a holistic dynamic segmentation network with scale-aware knowledge, Omni-Seg, that segments multiple tissue types at multiple resolutions using partially labeled images. The dynamic neural network based design with a scale-aware controller and the semi-supervised consistency regularization of pseudo-label learning achieves superior segmentation performance by modeling spatial correlations and consistency between different tissue types. The propose Omni-Seg method provides a generalizable solution for multi-scale multi-label segmentation in digital pathology, so as to ultimately leverage the quantitative clinical practice and research for various kidney diseases.



\section*{Acknowledgment}
This work was supported in part by NIH NIDDK DK56942(ABF).

\ifCLASSOPTIONcaptionsoff
  \newpage
\fi



\bibliographystyle{IEEEtran}
\bibliography{main}
%







\end{document}